\documentclass[conference, A4paper]{IEEEtran} 

\usepackage{amsmath}
\usepackage{amssymb}
\usepackage{amsfonts}
\usepackage{amsthm}
\usepackage{bm}

\usepackage{pdfcomment}

\usepackage[svgnames]{xcolor}
\usepackage{subfigure}
\usepackage{tikz}

\usepackage{acronym}

\usepackage[numbers,sort&compress]{natbib}

\usepackage{hyperref}

\begin{document}
\bstctlcite{BSTcontrol}


\title{Outage and Capacity Comparisons For Ground Relaying Systems Using Stochastic Geometry}

\author{\IEEEauthorblockN{Spyridon Evangelatos, Aris L. Moustakas  
   and~Andreas Polydoros
\thanks{This work was funded by the research program 'CROWN', through the Operational Program 'Education and Lifelong Learning 2007-2013' of NSRF, which has been co-financed by EU and Greek national funds.}}
\IEEEauthorblockA{ Department of Physics, University of Athens}
}


\maketitle

\begin{abstract}
Concurrent cooperative transmission for relaying purposes in mobile communication networks is relevant in current institutional systems with limited infrastructure, and and may be viewed as a potential range-extension mechanism for future commercial networks, including vehicular autonomous networking. The complexity of the overall system has encouraged certain abstractions at the physical layer which are critically analyzed in the present paper. We show via analytic stochastic-geometry tools that the receiver structure plays a crucial role in the outage behavior of the relays, particularly for realistic flooding protocols.  This approach aims to help understand the cross-layer aspects of such networks.
\end{abstract}


\section{Introduction}
\label{sec:Introduction}
Concurrent Cooperative Transmission (CCT) has been viewed as a potentially useful scheme in wireless networks where traditional networking infrastructure is either sparse (hence, requiring additional coverage-extension mechanisms), deficient, damaged or completely absent, as in certain institutional and military 
missions~\cite{Scaglione2003_OLA, Scaglione2006_OLA, Kailas2009_AOLA, Ramanathan2005_CRN, Jung2014_MPOLA, Halford2010_BRN0, Polydoros2014_PPDR}. Furthermore, a recent version of CCT for potential commercial use appears in~\cite{Ferrari2012_LowPowerWirelessBus}. The fundamental premise behind CCT is that  radio nodes in the network can be engaged to help transport information between a specific transmitter (Tx)-receiver (Rx) pair, or in a multicast-broadcast mode, whenever direct Tx-Rx connection is not of adequate quality. The assumption behind this premise is that the produced benefits of CCT are significant in terms of the usual figures of merit such as reliability, coverage, total network throughput, and so on, so as to justify the additional complexity required versus standard single-node transmission per relayed path. This is because CCT schemes imply a strong doze of inherent diversity: if a particular link fails due to the frailties of ground propagation, other concurrent links can provide the information to the intended receiver.

The present paper discusses the trade-offs and physical-level considerations that underlie CCT and examines the impact of signal-design choices (e.g., certain flavors of cooperation), network-geometry and propagation parameters on the achievable performance. We aim to quantify the gains in reception signal strength from multiple, randomly located relay nodes in a number of different settings and receiver architectures.  We assume a noise- and received-power-limited setting and thus focus on the statistics of signal-to-noise-ratio -- SNR from multiple concurrently transmitting sources. To gain insight we address a rather limited scenario, namely that of a spatially-enhanced, single-message, single-hop cooperative transmission towards a given  receiver. This can be considered as a building block for several multihop scenarios, such as sensor networks, ground repeaters extending a satellite's broadband transmission, military assets enhancing the integrity of a message in adverse ground environments, etc.
To make analytic progress, we assume that the locations of the transmitters relaying the same message follow a Point Poisson Process (PPP). Recent evidence \cite{Andrews2013_HETNET} suggests that PPP in an ad hoc network of mobile nodes, is a good model~\cite{Gastpar:2002:RelayCase}. 

\section{Communication-network model}\label{Sec:ComNet}
\begin{figure}[!t]
\centering
\includegraphics[scale=0.65]{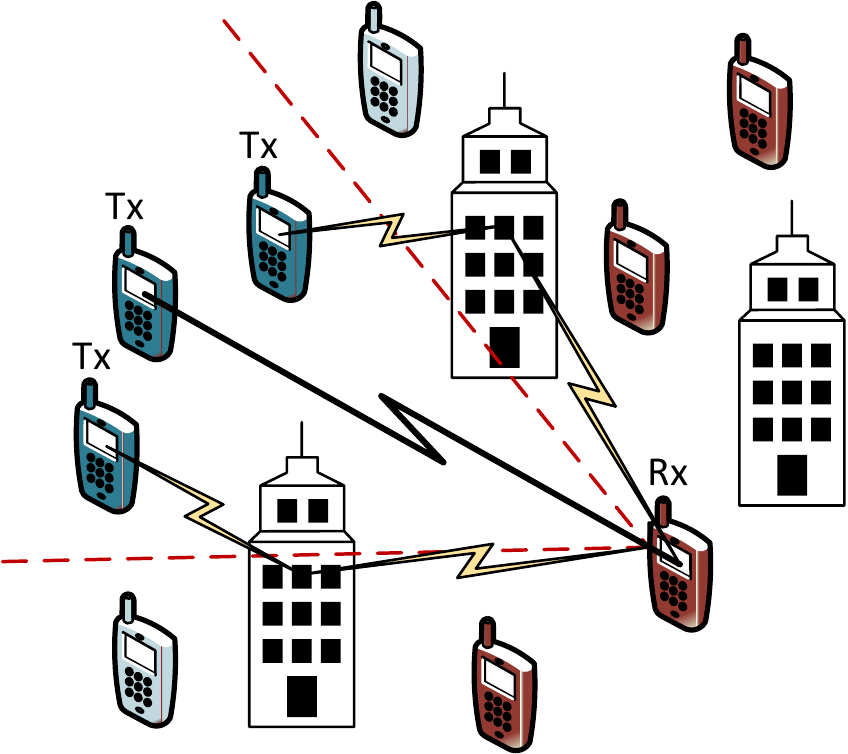}
\caption{An example of an ad hoc network where $3$ transmitters (Tx) relay  their signals to a single receiver (Rx). The transmitters exist in a slice of of width $\phi_{0}$ and their signals can be either in Line-of-Sight or reflected from the buildings. The cone geometry of transmitters mimics the fact that the destination is to the right of the Rx while the node of origin of the relay is to the left.}\label{fig:MANEt}
\end{figure}

To make specific quantitative progress, we start by defining a simplified model of the network. We envision a network with nodes distributed randomly in location and focus on a particular intermediate (or final) node. This receiver, located at the origin, receives signal from a set of
transmitters located stochastically with constant spatial density $\lambda$ over a cone of angular width $\phi_0\in[0,2\pi]$ (see Fig. \ref{fig:MANEt}). Any potentially interfering stream is assumed remote enough in space as to not add significant energy in the noise term. The adopted model does not incorporate spatial-interference effects. This is a valid simplification in at least two scenarios: one is when a chosen node acts as a scheduler, yielding the whole network to one source at a time, multiplexing them in time frames as per the sources' requests. This type of scheduling avoids interference altogether but is patently inefficient for non-broadcast traffic. A more efficient scheme would re-use space via the concept of Controlled Barrage regions. This scheme implies more sophisticated allocation of space to individual traffic streams of proper width .
Within each such region, interference from other streams would be minimal. The full theory for such spatial reuse is still a research topic and is not addressed here.

Three different concurrent-transmission (and therefore, respective reception) schemes of varying complexity and practicality will be analyzed here. The first, denoted as ``coherent'' below, is an idealized, upper-bounding case whereby each transmitter possesses a distinct orthogonal channel to itself for its message. The corresponding receiver can match-filter individually these dimensions and thus coherently demodulate and add all these contributions from all transmitters. In practical terms, this would necessitate mapping the independent transmissions onto orthogonal media (such as time slots, frequency bins or spreading codes) and subsequently collecting all the individual powers of all these transmissions at the receiver. Power collection would be meant both across orthogonal dimensions as well as across the multipath taps (ISI) for each dimension, utilizing either a RAKE receiver (for spreading schemes) or optimal maximum-likelihood sequence estimation for ISI. 

The second receiver scenario, denoted as ``incoherent'', is more practical, in which the signals from the individual transmitters arrive without phase coordination at each tap, namely, their voltage vectors add incoherently per tap, limiting the potential diversity gains of the signal. In an effort to combine these two transceiver models, we also consider the case, denoted as ``random'', where the transmitters have a finite number of orthogonal channels at their disposal with each transmitter using one such code. For simplicity in coordination (but a corresponding hit in the performance), we assume that each transmitter uses a randomly selected code. For a given instantiation of the Poisson point process denoted by $\bm{\omega}$ with $K(\bm{\omega})$ transmitters present, the corresponding received SNR can be expressed as
\begin{equation}\label{eq:SNR_c}
\mathrm{SNR}_{coh} = \sum_{k=1}^{K(\bm{\omega})} g_k f_k p_k \sum_{d=1}^{D} a_{d}|h_{kd}|^{2}
\end{equation}
In the above equation, $g_k=\ell_0/r_k^\alpha$ is the pathloss coefficient for the $k$th transmitter at distance $r_{k}$, with $\alpha$ being the pathloss exponent and $\ell_0$ the pathloss constant (see Table \ref{tab:ParamSett}). $f_k$ corresponds to the shadowfading fluctuations, assumed to be lognormal with $\mathbb{E}[f_k]=1$ and $\log_{10}f_k$ being a normal variable with variance $\sigma^{2}$. $p_{k}$ is the transmit power, which we normalize by the receive noise power (hence the ``$\mathrm{SNR}$\rq{}\rq{} term in~\eqref{eq:SNR_c} thus $p_k=P_k/(BW N_{0})$, where $P_k$ is the power of transmitter $k$, $BW$ is the bandwidth of the signal and $N_0$ is the noise spectral density. In~\eqref{eq:SNR_c} we model the frequency selectivity by assuming that there are $D$ distinguishable multipath components, with relative power strengths $a_d$, such that $\sum_d a_d=1$, possessing an exponential decay profile, as experimentally demonstrated in~\cite{Pedersen2000_SpaceTimeChannelModel}. As mentioned, the receiver match-filters the respective complex gains at each tap and then adds this matched power over all taps and all dimensions. This corresponds to the most optimistic collection of diversity, hence termed ``idealized\rq{}\rq{}. Moving to the incoherent case, the SNR is given by
\begin{equation}\label{eq:SNR_inc}
\mathrm{SNR}_{inc} = \sum_{d=1}^D a_d \left|\sum_{k=1}^{K(\bm{\omega})} g^{\frac{1}{2}}_k f^{\frac{1}{2}}_k p^{\frac{1}{2}}_k h_{kd}\right|^{2}
\end{equation}
The lack of phase coherence between transmitters in this case implies minimal spatial coordination and thus appears amenable to pragmatic deployment.  Since there is still a (theoretically) unbounded number of contributing transmitters, one might expect a finite diversity behavior for small SNR, even in the absence of multipath ($D = 1$). However, as we shall see in Section~\ref{SubSec:IncRec}, this will not be the case. Indeed, for small SNR the outage behavior is identical to that of a single transmitter! Finally, the SNR for the third, ``random'' case is
\begin{equation}\label{eq:SNR_rand}
\mathrm{SNR}_{rand} = \sum_{d=1}^Q \left|\sum_{k=1}^{K_q(\bm{\omega})} g^{\frac{1}{2}}_k f^{\frac{1}{2}}_k p^{\frac{1}{2}}_k h_{kd}\right|^{2}
\end{equation}
Now, a different random subset of nodes transmits in each dimension, with corresponding density $\lambda/Q$. To keep the analysis simple in this case, we assume flat fading $D=1$. The results for this case are discussed in Section~\ref{sec:RandomAllocationChannels}.
\section{Reception Statistics}\label{Sec:StatRec}

\subsection{Coherent Reception}\label{SubSec:CohRec}

In this section we calculate the statistics of coherent detection as defined in \eqref{eq:SNR_c}. Some of the expressions obtained here are known for pathloss exponent $\alpha=4$, but not for general $\alpha$. Via standard transform techniques (see, for example,~\cite[Ch.\ 3.2]{Haenggi2008_InterferenceLargeWirelessNets}), we compute the Laplace transform ${\cal L}(p)$ of $\mathrm{SNR}_{coh}$:
\begin{eqnarray} \label{eq:LapXform_scoh}
\log {\cal L}(u) = \log \mathbb{E}\left[e^{-u\mathrm{SNR}_{coh}}\right] = - \overline{\lambda}  u^{\frac{2}{\alpha}}
\end{eqnarray}
with the constant $\overline{\lambda}$ being
\begin{eqnarray}\label{eq:Lambda_bar_def}
\overline{\lambda} =\lambda \phi_0 \Gamma\left(1-\frac{2}{\alpha}\right)  \left(\frac{P\ell_{0}}{N_{0}BW}\right)^{\frac{2}{\alpha}} \mathbb{E}\left[z^{\frac{2}{\alpha}}\right] \mathbb{E}\left[f^{\frac{2}{\alpha}}\right]
\end{eqnarray}
where $\Gamma(x)$ is the Euler $\Gamma$ function and the expectation over the random variable $z = \sum_d a_{d}|h_{d}|^{2}$, after averaging over the fast-fading complex Rayleigh components $h_d$ from a fixed node can be expressed as \cite{Moustakas2002_MISO1}
\begin{equation}\label{eq:<z^2/a_2}
 \mathbb{E}[z^{\frac{2}{\alpha}}] = \Gamma\left(1+\frac{2}{\alpha}\right) \sum_{d=1}^{D} \frac{a_{d}^{\frac{2}{\alpha}}}{\prod_{\mu\neq d}\left(1-\frac{a_{\mu}}{a_{d}}\right)}
\end{equation}
The expectation over the lognormal shadow fading component $f$ gives $ \mathbb{E}[f^{\frac{2}{\alpha}}]=\exp\left[\frac{\sigma^2}{\alpha}\left(\frac{2}{\alpha}-1\right)\right]$ leading to
\begin{equation}\label{eq:Lambda_bar}
\overline{\lambda} = \frac{2\pi \lambda \phi_0 }{\alpha\sin\frac{2\pi}{\alpha}}\left(\frac{P\ell_{0}}{N_{0}W}\right)^{\frac{2}{\alpha}}
e^{^{\frac{2\sigma^2}{\alpha^2}-\frac{\sigma^2}{\alpha}}} \sum_{d=1}^{D} A_d a_{d}^{^{\frac{2}{\alpha}}}
\end{equation}
where
\begin{equation}
A_{d} = \prod_{k=1\neq d}^{D}\left(1-\frac{a_{k}}{a_{d}}\right)^{-1}
\end{equation}
To obtain $f_{coh}(s)$,  the distribution of $\mathrm{SNR}_{coh}$ we need to invert  the Laplace transform in \eqref{eq:LapXform_scoh}. For general $\alpha>2$ this cannot be performed exactly.
Nevertheless, the tails for the distribution for small $s\ll \overline{\lambda}^{\alpha/2}$ can be obtained by applying the saddle point method (see \cite{Carrier_Krook_Pearson_Book_complex_analysis}), which yields
\begin{equation}\label{eq:InvLapX_s1_sliceL_gen_alpha}
f_{coh}(s) \approx G_{D,\alpha}\, s^{-\frac{\alpha-1}{\alpha-2}}\exp\left[-\frac{\alpha-2}{\alpha} \left(\frac{2\left(\overline{\lambda}\right)^{\frac{\alpha}{2}}}{\alpha s}\right)^{\frac{2}{\alpha-2}}\right]
\end{equation}
where
\begin{equation}
G_{D,\alpha} = \frac{\left(\frac{2}{\alpha}\right)^{\frac{1}{\alpha-2}} }{\sqrt{\pi(\alpha-2)}} \left(\overline{\lambda}\right)^{\frac{\alpha}{2(\alpha-2)}}
\end{equation}
For $\alpha=4$ the above equation is exact, which generalizes the result of \cite{Haenggi2008_InterferenceLargeWirelessNets} to multiple paths and the sector geometry. We see that the effect of restricting the reception to a finite sector in space reduces the effective density of the transmitters. Also note that the effect of multiple paths is to increase the effective density of transmitters, or equivalently the effective diversity. The most important feature of this result is that the probability of very small received signal $s\ll 1$ becomes extremely small. In fact, the key benefit of coherent reception is that the probability density at $s = 0$ is zero, in a nonanalytic way. 

\subsection{Incoherent Reception}\label{SubSec:IncRec}
We now analyze the case of incoherent reception, with $\mathrm{SNR}_{inc}$ given in \eqref{eq:SNR_inc}.  For the general case of an $D$-tap channel \eqref{eq:SNR_inc} can be written as
\begin{equation}\label{eq:signal_coherent_def2}
\mathrm{SNR}_{inc}  = \sum_{d=1}^{D} a_{d} \bm{h}^{\dagger}_{d} \bm{\Delta} \bm{h}_{d}
\end{equation}
where $\bm{h}_{d}$ is the complex vector of the fading coefficients $h_{id}$ for $i=1,\ldots,K(\bm{\omega})$ and $\bm{\Delta}$ is a rank one matrix with elements $\Delta_{ij}=\left(g_{i}f_{i}p_{i}g_{j}f_{j}p_{j}\right)^{\frac{1}{2}}$. Since the vector $\bm{h}_{d}$ is complex Gaussian it can be readily integrated out when calculating the Laplace transform. Thus for a fixed number of transmitter nodes the Laplace transform can be expressed as,
\begin{eqnarray} \label{eq:LapXform_s21}
{\cal L}(u) &=& \mathbb{E}\left[\prod_{d=1}^{D} \frac{1}{1+u a_d \sum_{k=1}^K g_k f_k p_k}\right] \\ \nonumber
&=& \prod_d \mathbb{E}\left[\int_0^\infty dq_{d}\, \exp\left[-q_{d} - u q_{d} a_d \sum_{k=1}^K g_{k} f_{k}p_{k} \right]\right]
\end{eqnarray}
As a result, after averaging over node numbers, positions, and shadow fading we obtain
\begin{equation} \label{eq:LapXform_s2_sliceL}
{\cal L}(u) =\prod_{d=1}^{D}  \int_{0}^{\infty} dq_{d} \exp\left[-\sum_{d=1}^D q_{d} - u^{\frac{2}{\alpha}}\,\widehat{\lambda} z(\bm{q})^{\frac{2}{\alpha}}\right]
\end{equation}
where $z(\bm{q}) = \sum_{d} a_{d} q_{d}$ and
\begin{equation}
 \widehat{\lambda} = \lambda \left(\frac{P\ell_{0}}{N_{0}W}\right)^{\frac{2}{\alpha}} \phi_0 \Gamma\left(1-\frac{2}{\alpha}\right) \exp\left[\frac{\sigma^2}{\alpha}\left(\frac{2}{\alpha}-1\right)\right]
\end{equation}
The above expression may be inverted exactly for the case of $\alpha=4$. To do this we first Laplace transform to obtain a similar expression with \eqref{eq:InvLapX_s1_sliceL_gen_alpha} and then integrate over $q_d$ to obtain
\begin{equation}\label{eq:InvLapX_s2_1slice}
f_{inc}(s) = \sum_{d=1}^{D}\frac{2A_{d}\,\widehat{\lambda}\,\sqrt{a_{d}}}{\left(4 s+a_{d}\widehat{\lambda}^2\right)^{\frac{3}{2}}}
\end{equation}
The above formula, when compared to the corresponding one of coherent reception in~\eqref{eq:InvLapX_s1_sliceL_gen_alpha} shows that while the large-s behavior is asymptotically the same, namely $f(s)\sim s^{-3/2}$,
its small-$s$ behavior is starkly different. 

In particular, for $D=1$ we find a finite probability density for $s=0$. This is due to the fact that due to the random phases between the different complex channel coefficients any combination of pathloss terms may become arbitrarily small. In the case of $D>1$, the small $s$ behavior of $f_{inc}(s)$ can be obtained for general $\alpha$ by using saddle-point analysis of the inverse Laplace transform of \eqref{eq:LapXform_s2_sliceL}, so that
\begin{equation}\label{eq:InvLapX_s2_Lslice_smallx}
f_{inc}(s) \approx B_{D,\alpha} \frac{s^{D-1}}{{\widehat{\lambda}}^{\frac{\alpha D}{2}} }
\end{equation}
where $B_{D,\alpha}$ is a constant given by
\begin{equation}
B_{D,\alpha} = \frac{\Gamma\left(\frac{(\alpha-2)D+1}{2}\right)  \left(\frac{\alpha}{2}\right)^{D+1}\left(\frac{\alpha}{\alpha-2}\right)^{\frac{(\alpha-2)D-1}{2}} }{\Gamma(D)\sqrt{\pi (\alpha-2)} \prod\limits_{d=1}^{D} a_{d}}
\end{equation}
Here, the presence of independently-contributing multipath taps reduces the outage probability (small-$s$ behavior) and alleviates the incoherence problem.
\begin{figure}[!t]
\centering
\includegraphics[scale=0.35]{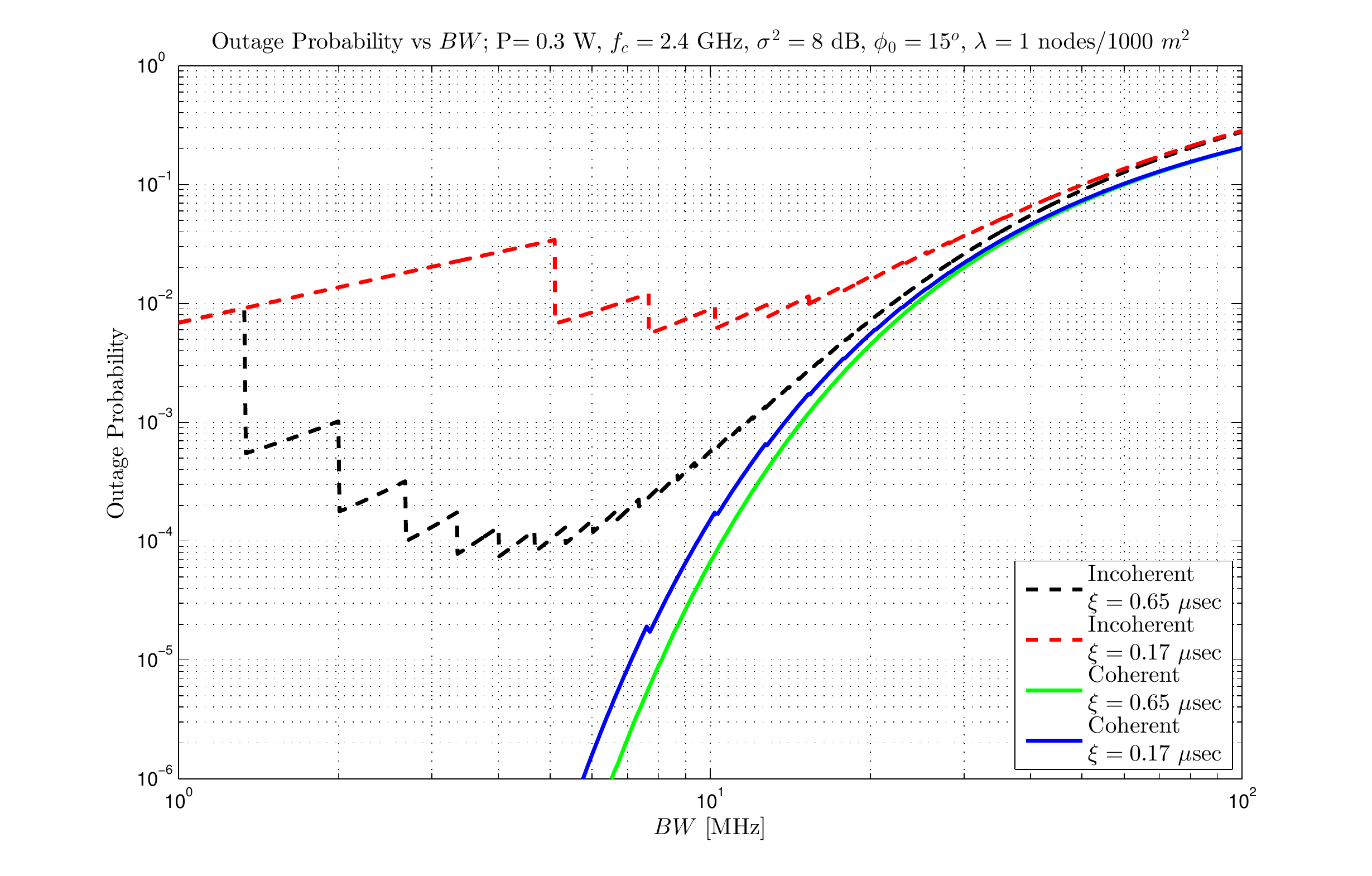}
\caption{Outage probability for coherent and incoherent reception as a function of bandwidth for two different values of $\xi$ and fixed node density $\lambda=1$ node/$1000\, m^2$ and SNR$=0$dB.}\label{fig:OutProbTap}
\end{figure}

\subsection{Random Allocation of Orthogonal Channels}
\label{sec:RandomAllocationChannels}

We now analyze the case of reception from a finite number of orthogonal channels, with $\mathrm{SNR}_{rand}$ given in \eqref{eq:SNR_rand}.  For simplicity we only treat the single tap $D=1$ case. Using identical arguments as above, we express the Laplace transform of~\eqref{eq:SNR_rand}
\begin{eqnarray} \label{eq:LapXform_s2_sliceQ}
{\cal L}(u) =\prod_{\ell=1}^Q \left\{\int_0^\infty dq_\ell \exp\left[-q_{\ell} - u^{\frac{2}{\alpha}}\frac{\tilde{\lambda}}{Q} q_{\ell}^{\frac{2}{\alpha}}\right] \right\}
\end{eqnarray}
When $Q>1$ $f(s)$ cannot be written in closed form. However,  its small $s$ tails can be obtained using saddle-point analysis of the inverse Laplace transform of \eqref{eq:LapXform_s2_sliceL}, so that
\begin{equation}\label{eq:pdf_rand_smallx}
f_{rand}(s) \approx C_{Q,\alpha} \frac{s^{Q-1}}{ {\widehat{\lambda}}^{\frac{\alpha Q}{2} } }
\end{equation}
where $C_{Q,\alpha}$ is a constant given by
\begin{eqnarray}
C_{Q,\alpha} = \frac{\Gamma\left(\frac{(\alpha-2)Q+1}{2}\right)\Gamma\left(\frac{\alpha}{2}\right)^{Q} \left(\frac{\alpha}{2}\right)^{2Q} \left(\frac{\alpha-2}{\alpha}\right)^{\frac{1-(\alpha-2)Q}{2}}  }{\sqrt{\pi (\alpha-2)}\Gamma\left(\frac{\alpha Q}{2}\right) Q^{-\frac{\alpha Q}{2}} }
\end{eqnarray}
Hence, we see that the number of orthogonal paths here saves the day and reduces the outage for small $s$.

\section{Performance Measures}
\subsection{Outage Probability}
In this section we evaluate the outage distribution, to establish performance metric for the above schemes. Starting with coherent detection, we integrate~\eqref{eq:InvLapX_s1_sliceL_gen_alpha}  for $\alpha=4$  to obtain
\begin{eqnarray}\label{eqn:OutCohRec}
\mathbb{P}\left(\mathrm{SNR}_{coh}<s\right) = \mathbb{P}_{out}^{coh}(s) = 2\,\mathbb{Q}\left(\frac{\overline{\lambda}}{\sqrt{2s} }\right)
\end{eqnarray}
Similarly, the outage probability for incoherent reception for $\alpha=4$ can be obtained from \eqref{eq:InvLapX_s2_1slice} to be
\begin{eqnarray}\label{eqn:OutIncRec}
\mathbb{P}_{out}^{inc}(s)  = 1-\sum_{d=1}^{D}\frac{A_{d}\sqrt{a_{d}}\,\widehat{\lambda}}{\sqrt{4s+a_{d}\widehat{\lambda}^{2}}}
\end{eqnarray}
For general $\alpha\neq 4$ the small $s$ behavior of the $\mathbb{P}_{out}^{inc}(s) $ can be obtained from~\eqref{eq:InvLapX_s2_Lslice_smallx} to be
\begin{equation}
\mathbb{P}_{out}^{inc}(s) \approx \frac{B_{D,\alpha}}{D}\left(\frac{s}{  {\widehat{\lambda}}^{\frac{\alpha}{2}} }\right)^D
\end{equation}

Finally, for small $s$ the outage probability for reception from $Q$ orthogonal codes is approximated by
\begin{equation}
\mathbb{P}_{out}^{rand}(s) \approx \frac{C_{Q,\alpha}}{Q}\left(\frac{s}{  {\widehat{\lambda}}^{\frac{\alpha}{2}} }\right)^Q
\end{equation}

\subsection{Capacity of Relaying Systems}
We can express the ergodic capacity for the incoherent reception $\mathcal{C}_{inc}(\widehat{\lambda}) $ and $\alpha=4$
 as follows: If $a_{d}\widehat{\lambda}^{2} < 4$, then
\begin{equation}
\mathcal{C}_{inc}=\sum_{d=1}^{D}\frac{\sqrt{a_{d}}\widehat{\lambda}A_{d}}{\sqrt{4 - a_d\widehat{\lambda}^{2}}} \text{arccot}\left(\frac{a_{d}\widehat{\lambda}}{\sqrt{4 - a_{d}\widehat{\lambda}^{2}}}\right)
\end{equation}
On the other hand, if $a_{d}\widehat{\lambda}^{2} > 4$, then
\begin{equation}
\mathcal{C}_{inc}=\sum_{d=1}^{D}\frac{2\sqrt{a_{d}}\widehat{\lambda}A_{d}}{\sqrt{a_{d}\widehat{\lambda}^{2} - 4}}\ln\left(\frac{\sqrt{a_{d}}\widehat{\lambda}  + \sqrt{a_{d}\widehat{\lambda}^{2} - 4} }{2}\right)
\end{equation}
For coherent reception, the general capacity formula is unobtainable; however, for $\alpha=4$,
\begin{equation}
\mathcal{C}_{coh} 
=\pi \text{Erfi}\left(\frac{\overline{\lambda}}{2}\right) - \frac{\overline{\lambda}^2}{2} {}_2F_2\left((1,1);\left(\frac{3}{2},2\right);\frac{\left(\overline{\lambda}\right)^2}{4}\right)
\end{equation}
where $ _2F_2(\cdot)$ is the Hypergeometric function~\cite{Abramowitz1964_Handbook}.
\begin{figure}[!t]
\centering
\includegraphics[scale=0.35]{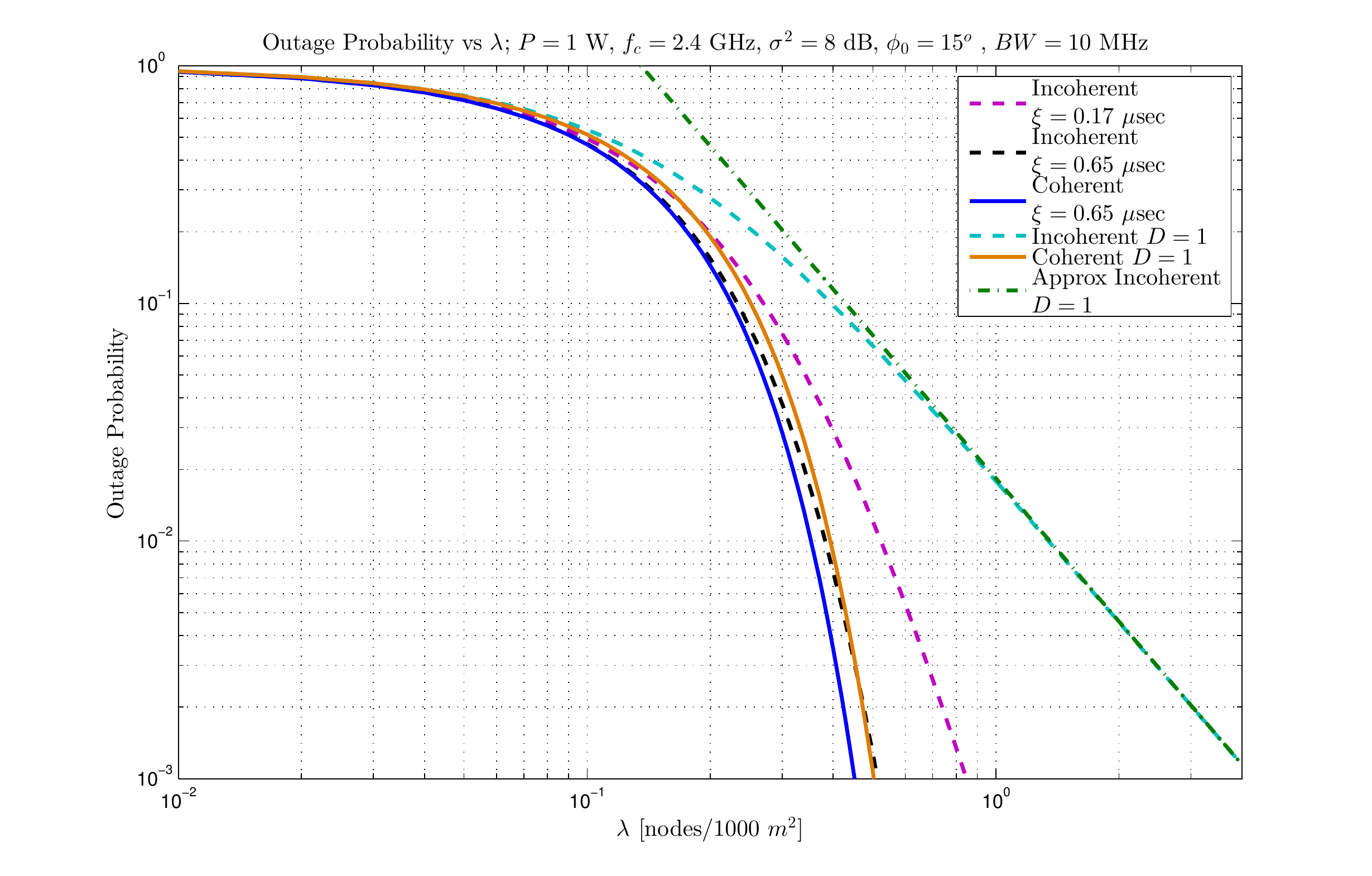}
\caption{Outage probability (and its approximations) as a function of the node density $\lambda$ for fixed SNR$=0$ dB and bandwidth $BW=10$ MHz along with the cases where $D=1$}\label{fig:OutProbDen1}
\end{figure}\hfill
\begin{table}[!t]
\centering
\begin{tabular}{|p{5cm}|p{3cm}|}
\hline
{\bf Parameter} & {\bf Value} \\ \hline
Carrier Frequencey & $f_{c}=2.4$ GHZ\\    \hline
Delay Spread Values \cite{Calcev2004_3GPP_SCM}& $\xi=0.17\,\mu s$  and $\xi=0.65\,\mu s$ \\    \hline
Lognormal Shadowing Standard Deviation & $8$ dB \\    \hline
Bandwidth & $BW=10$ MHz\\    \hline
Noise Power & $N_0=-97.8$ dBm \\    \hline
Average Pathloss Attenuation at $\overline{x}=25$m & $\ell_{0} = -93$ dB \\    \hline
Pathloss Exponent & $\alpha=4$ \\    \hline
\end{tabular}\caption{Parameter settings for the numerical results.}\label{tab:ParamSett}
\end{table}
\section{Numerical Results}\label{Sec:SimRes}
\begin{figure}[!t]
\centering
\includegraphics[scale=0.35]{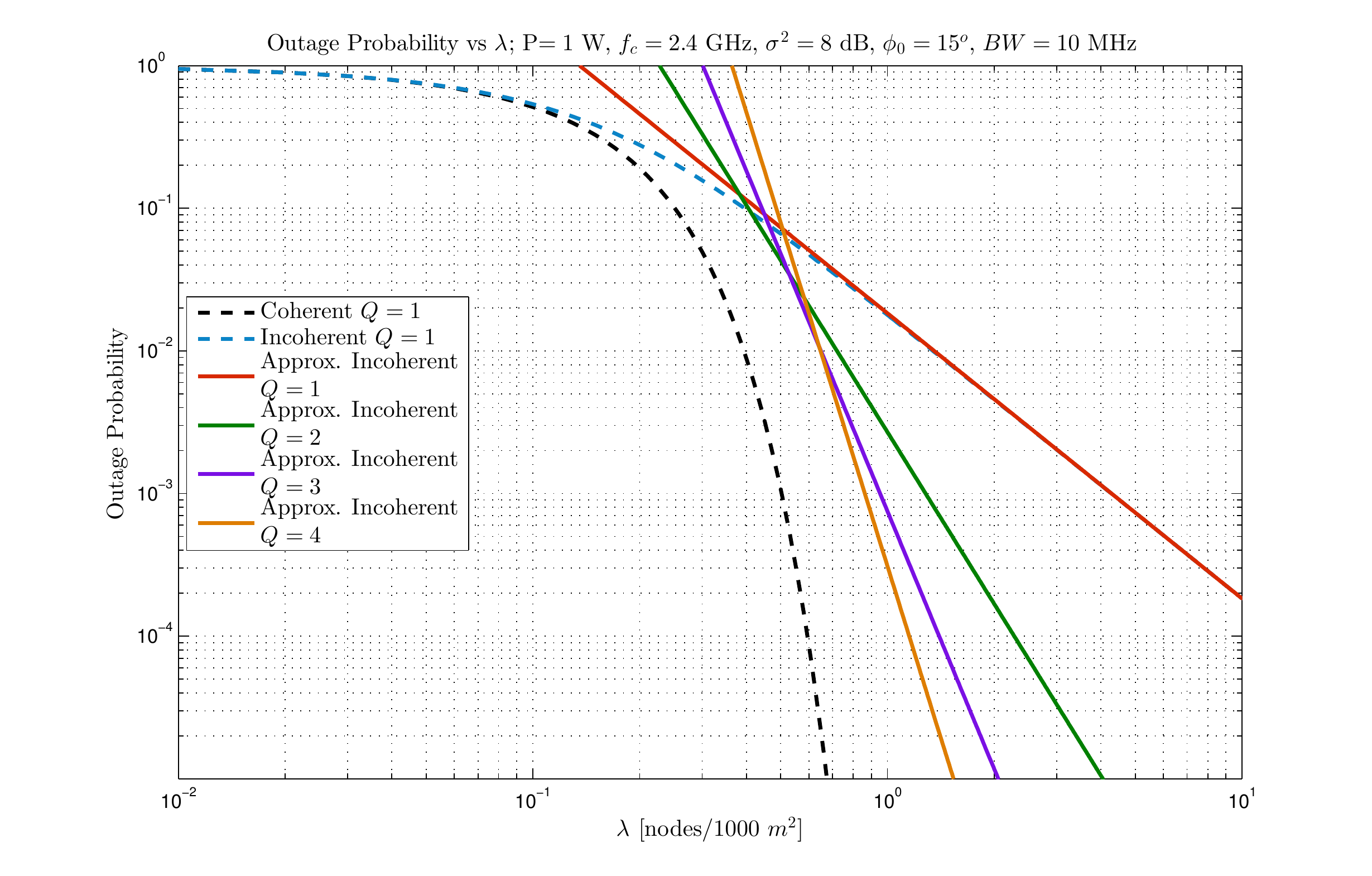}
\caption{Outage probability as a function of $\lambda$ along with its approximations for the incoherent reception, for $BW=10$ MHZ.}\label{fig:OutProbQPaths}
\end{figure}

\begin{figure}[!t]
\centering
\includegraphics[scale=0.35]{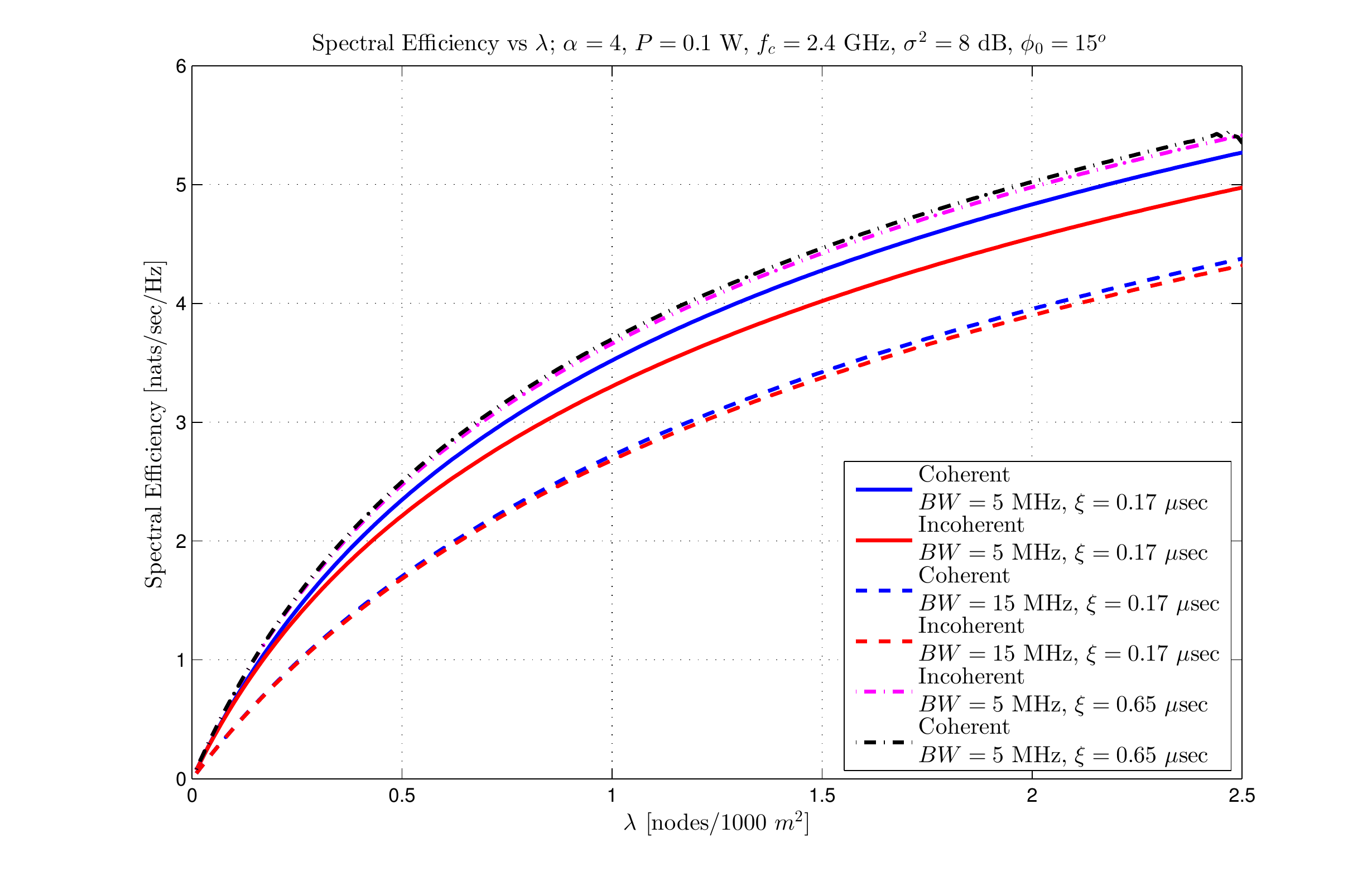}
\caption{Spectral efficiency as a function of node density $\lambda$ for two different values of bandwidth $BW$ and delay spread $\xi$.}\label{fig:OutProbDen}
\end{figure}

We now draw certain quantitative conclusions from the previous results. The chosen parameter values are summarized in Table~\ref{tab:ParamSett}. $a_d$, the relative strength of the receiver power in tap $d$  for a given bandwidth BW and delay spread  $\xi$ is expressed as
\begin{equation}\label{eqn:RelSigStr}
a_{d}= n_{0}\left(\exp\left[-\frac{(d-1)}{(BW\,\xi)}\right] -\exp\left[-\frac{d}{(BW\,\xi)}\right]\right)
\end{equation}
where $n_0$ is a normalization constant so that $\sum_d a_d=1$. The number of taps $D$ is determined from the condition that $90\%$ of the total power is captured. Fig.~\ref{fig:OutProbDen1} depicts the outage probability versus $\lambda$. This is equivalent to plotting versus SNR, since the outage probability depends on the parameter $\lambda P^{2/\alpha}$, thus allowing for a tradeoff between node density and power. While the outage for coherent reception is largely independent of the delay spread (and therefore $D$), the outage probability of incoherent reception depends strongly on $\xi$. Indeed for $\xi=0.65~\mu sec$ corresponding to urban environments~\cite{Calcev2004_3GPP_SCM} the outage is essentially identical to that of coherent reception. Fig.~\ref{fig:OutProbTap} depicts the outage probability versus bandwidth. The dependence on bandwidth is two-fold. First, for increasing bandwidth the number of taps increases hence reducing the outage. However at the same time the noise power increases, thereby increasing the outage. Thus for a given delay spread there is a  optimum bandwidth. In any case, the outage for the incoherent case, although higher than that of the coherent one, is still relatively low. Fig.~\ref{fig:OutProbQPaths} compares the outage behavior of all three reception schemes discussed here. We see that indeed the case with finite number of orthogonal channels $Q$ is inbetween the two other cases (coherent and incoherent). For moderate values of $Q$ the slopes of the outage become close to that of coherent reception, indicating that not too many orthogonal channels can provide enough diversity to the system. Finally, in Fig.~\ref{fig:OutProbDen} we plot the ergodic spectral efficiency versus the density for various values of delay spread and bandwidth. We find that increasing delay spread increases the spectral efficiency, while increasing bandwidth has the opposite effect.

\section{Conclusion}\label{Sec:Rem}
We have introduced fairly detailed models of reception in stochastic-geometry relaying networks and assessed their impact analytically in the outage regime. Pragmatic protocols of the stateless type tend to suffer from the phase-noncoherent combination of same-tap simultaneous receptions, which implies the need to counter that with other means. Rich delay-spread multipath, if present and properly collected, is shown to have an ameliorating effect on outage. The coherent power-collecting model is shown to be superior in outage. However, the latter implies protocol-level coordination as well as signal expansion in the resource dimension (not required in the noncoherent model), with a concomitant loss of throughput at the cooperative-link level. A more thorough study of end-to-end throughput would quantify the associated tradeoffs.
\section*{Acknowledgments}
This work was funded by the research program `CROWN\rq{}, through the Operational Program `Education and Lifelong Learning 2007-2013\rq{} of NSRF, which has been co-financed by EU and Greek national funds.

\footnotesize
\bibliographystyle{IEEEtran}
\bibliography{IEEEabrv,brn_bib,C:/Users/ARISLM/ALMDocuments/Dropbox/Work/CurrentWork/bibliography/wireless}
\end{document}